\documentclass[a4paper,10pt]{article}
\usepackage[latin1]{inputenc}
\usepackage[T1]{fontenc}
\usepackage[normalem]{ulem}
\usepackage{verbatim}
\usepackage{amssymb}
\usepackage{vmargin}
\usepackage{graphicx}
\setmarginsrb{2.5cm}{2.5cm}{2.5cm}{2.5cm}{0cm}{0cm}{0cm}{1cm}

\title{Thermocapillary flows and interface deformations produced by localized laser heating in confined environment}

\author{By\\ Hamza Chra\"ibi and Jean-Pierre Delville\\
\small Univ. Bordeaux, LOMA, UMR 5798, F-33400 Talence, France.\\
\small CNRS, LOMA, UMR 5798, F-33400 Talence, France.}

\begin{document}
\maketitle
\begin{abstract}
 The  deformation of a fluid-fluid interface due to the thermocapillary stress induced by a continuous Gaussian laser wave is investigated analytically.
 We show that the direction of deformation of the liquid interface strongly depends on the viscosities and the thicknesses
of the involved liquid layers. We first investigate the case of an interface separating two different liquid layers while a second part is dedicated 
to a thin film squeezed by two external layers of same liquid. These results are predictive for applications fields where localized thermocapillary
 stresses are used to produce flows or to deform interfaces in presence of confinement, such as optofluidics.
\end{abstract}

\section{Introduction}
  When a Gaussian laser beam heats the interface separating two superimposed optically absorbing liquids, the inhomogeneous heating induces a local variation
 of the interfacial tension which in turn generates tangential shear stresses and sets up hydrodynamic flows known as thermocapillary or Marangoni flows \cite{schatz01}.
 These laser-induced thermocapillary flows were studied theoretically, experimentally and numerically in various cases, with a particular emphasis
 to the free-surface configuration. In the eighties, Loulergue et al. \cite{loulergue81} investigated theoretically the heating of a flat fluid interface by a spatial
 sine modulation of an infrared laser beam to produce an infrared image converter. 
At the same time, Viznyuk et al. \cite{viznyuk88},
 investigated the free-surface deformation of thin layers by Gaussian laser beams and applied their theoretical results to beam
 shaping using what they called optocapillarity.
Beyond optical applications, Longtin et al. \cite{longtin99} investigated time dependent
 behaviors of free-surfaces by numerical simulation and intended a comparison between predictions for laser pulse heating
 with experiments and scaling analyses, the goal being to get new insights in laser melting and welding \cite{mackwood05}.
 One can also cite the theoretical analysis of Rivas \cite{rivas91} who studied the effect of the viscous stress on the interface deformation
and the numerical work of Marchuk \cite{marchuk09} who investigated the effect of convection.\\
 We know that the temperature variation of the surface tension is usually negative for classical liquids \cite{escobedo98}.
 Therefore, fluids are almost always pushed toward coldest regions and the previous experiments showed that laser heating
 produces beam centered dimple at the interface. This led Bezuglyi et al. \cite{bezuglyi01} to consider very thin layers
 and investigate the possibility to optically control film rupture and hole formation in order to understand new mechanisms
 for varnish dewetting, wetting \cite{nagy08} and spreading \cite{garnier03} of liquid drops and microfilms over solid substrates.
	However, contrary to usual expectations, Misev \cite{Mizev04} experimentally observed a curvature inversion
 of the interface deformation, i.e. a concave to convex transition, when decreasing the thickness of the fluid layer.
 Therefore, it appears that dimple formation does not represent a general rule when fluid confinement starts to play a role,
 as already supported by classical Marangoni experiments in thin cells \cite{vanhook97}. This transition from concave
 to convex interface deformation has been studied recently \cite{karlov05} but the generalization to two-liquid systems
 still deserves to be investigated. This is the goal of the present work.\\
	We theoretically investigate thermocapillary effects induced by a Gaussian laser beam that locally heats
 the interface separating two liquid layers or the interfaces of a thin film bounded by two liquid layers.
 Beyond the general description of flow patterns and interface deformations in many different situations,
 our investigation also provides new insights for digital optofluidics
 applications \cite{delville09},\cite{baroud10}, where light-actuated droplets are naturally confined and often close together, squeezing by the way thin liquid layers.\\
 Our study is structured as follows: the theoretical resolution of the flow and the interface deformation 
is detailed in section 2. Section 3 reports results and discussion on the dependence of the flow and the deformation to the viscosity and layer thicknesses ratios
 in two different configurations. We first consider an interface separating
 two different liquid layers and then generalize these results to the case of thin film separating two same liquid layers.\\
\section{Two-Fluid theoretical model}
\begin{figure}[h!!!]
\begin{center}
 \includegraphics[scale=0.5]{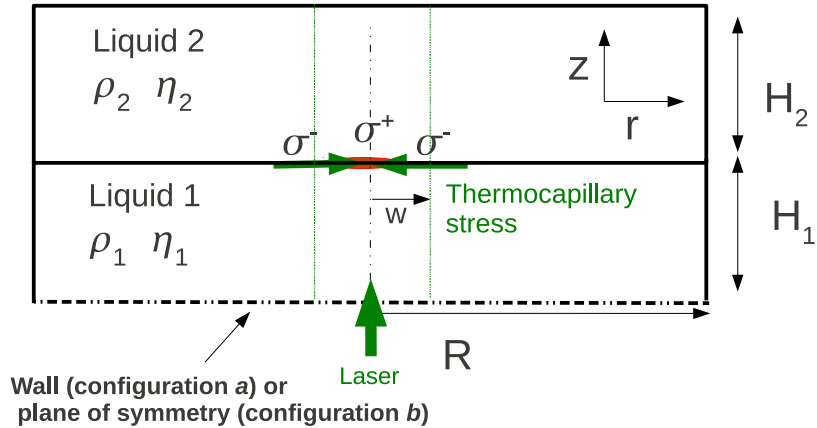}
\caption{Schematic representation of the liquid-liquid system heated by a laser beam. See text for other notations.
$\sigma^+$ indicated that the interfacial tension increases with temperature. In this configuration, we arbitrarily consider $\partial \sigma / \partial T>0$, a situation often encountered
 in microfluidics due to the presence of large amount of surfactant. }\label{sketch}
\end{center}
\end{figure}
 Let us consider two liquid layers separated by an interface initially flat, horizontal and at rest, crossed by a continuous Gaussian laser beam propagating perpendicularly 
to the interface (see figure \ref{sketch}).
 If at least one of the layers absorbs light, the interface will be locally heated by the laser.
 Physical properties of the liquids (denoted $1$ for the bottom liquid and $2$ for the top one) 
are their viscosities $\eta_1$, $\eta_2$ and densities $\rho_1$, $\rho_2$. 
The interfacial tension is denoted by $\sigma$.
 Liquids are enclosed in a cylindrical cell of radius $R>>w$, where $w$ is the radius of the temperature distribution 
due to laser heating.\\
Considering the axisymmetry, along the $z$ axis of the exciting beam, of the temperature distribution, cylindrical coordinates (${\bf e_r, e_\theta, e_z})$
with origin $O$ located at the intersection of the beam axis
with the initial flat interface are chosen for this study.  A point $\textbf{x}$ is thus referenced by the space
coordinates $(r,\theta,z)$.
Two different configurations are investigated. The first one (a) considers two liquid layers separated by an interface,
 and the second one (b) generalizes the two-layer situation to a symmetric system composed of three liquid layers, liquid $1$ being bounded on top and bottom by liquid $2$ (see figure \ref{sketch}).\\
\subsection{Heat equations}
The temperature distribution within the fluid layers due to the heating sources $\Upsilon_i$ obeys the heat equation 
\begin{equation}
\frac{\partial T_i}{\partial t}+({\bf u_i\cdot\nabla})T_i=D_{T_i}\nabla^2T_i+\Upsilon_i~~~~i=1,2\label{heat}
\end{equation}
${\bf u_i}$ and $D_{T_i}$ being the velocity and the thermal diffusivity of fluid $i$.\\
Assuming, without the loss of generality, that only fluid $1$ absorbs light, $\Upsilon_2=0$  and
\begin{equation}
\Upsilon_1=\frac{D_{T_1}}{\Lambda_1}a_1 I(r,z)
\end{equation}
where $\Lambda_1$ and $a_1$ are respectively the thermal conductivity and the optical absorption of fluid $i$ at the used optical wavelength.\\
Light intensity distribution is given by
\begin{equation}
I(r,z)=\frac{2P}{\pi \omega_0^2}\exp\left(\frac{-2r^2}{\omega_0^2}-a_1z\right)
\end{equation}
where $P$ is the beam power and $\omega_0$ the radial extension of the laser beam, also called beam waist. $\omega_0$ is always smaller than $w$ due to 
the non locality and the strong dependence in boundary conditions of temperature distributions.\\
The keypoint for producing thermocapillary flows is not the overheating itself but the amplitude of the temperature gradient. Laser heating is thus very appealing
 because it is extremely easy to produced weak amplitude overheating with large gradients.
Thus we can confidently consider a weak optical absorption in fluid $1$ such as $a_1z \leqslant a_1H_1 \ll 1$, and neglect the axial variation of light intensity and therefore
\begin{equation}
I(r)\simeq \frac{2P}{\pi \omega_0^2}\exp\left(\frac{-2r^2}{\omega_0^2}\right)
\end{equation}
The convective and unsteady terms of equation (\ref{heat}) can be neglected when considering that the thermal P\'eclet number is small 
($\displaystyle{Pe_T=\frac{u_0 w}{D_{T_i}} \ll 1}$, where $u_0$ is a characteristic velocity)
 and a characteristic heating time $w^2/D_{T_i}$ small compared to the viscous characteristic time $w^2/\nu_i$, where $\nu_i$ is the kinematic viscosity of fluid $i$.\\
An example of experiment fitting these assumptions can be found in a recent publication \cite{rsv12}.\\
Exact solutions of the heat equation using continuity of temperature and heat flux at the interface can be calculated \cite{rsv12}. However for the sake of simplicity we will consider a Gaussian distribution for $T_i(r)$ 
which is a reasonable approximation of the exact temperature field.\\
\subsection{Fluid equations of motion}
Both fluids obey the mass conservation and Stokes equations, and are coupled through stress balance
 in addition to the continuity of velocity at the interface described by its height $h(r)$ and denoted $S_I$. As flows are expected
to be viscous, the boundary value problem can be expressed as\\
\begin{equation}
\nabla . {\bf u_i} = 0~~;~~  i=1,2
\label{mass}
\end{equation}
\begin{equation}
{\bf 0} = -\nabla p_i + \eta_i\Delta {\bf u_i}~~;~~ i=1,2
\label{stokes}
\end{equation}
\begin{equation}
({\bf T_{1}}\cdot {\bf n}- {\bf T_{2}}\cdot {\bf n})=(\sigma \kappa (r) - (\rho_1-\rho_2)g z)
{\bf n}+\frac{\partial \sigma}{\partial s}{\bf t}~~;~~ {\bf x}~ \in~ S_I \label{stressjump}
\end{equation}
\begin{equation}
{\bf u_1} = {\bf u_2}~~;~~ {\bf x}~ \in~ S_I \label{cont_v}
\end{equation}
 $p_i$ is the pressure in each fluid and $\displaystyle{{\bf T_i}
= -p_{i} {\bf I} + \eta_i(\nabla {\bf u_i} + ^t\nabla {\bf u_i})}$  is the stress tensor corrected with a gravity term.
 The unit vector normal to the interface directed from fluid 1 to fluid 2 is denoted by $\bf{n}$,
 $s$ is the arc length on the interface, and ${\bf t}$ is the unit vector tangent to the interface such as ${\bf (t,e_\theta,n)}$ is orthonormal.\\
In Equation (\ref{stressjump}), $\displaystyle{\kappa(r)=\frac{1}{r}
\frac{d}{dr}\frac{r\frac{dh}{dr}}{\sqrt{1+{\frac{dh}{dr}}^2}}}$ is the double mean curvature of the
axisymmetric interface in cylindrical coordinates. $\displaystyle{\frac{\partial \sigma}{\partial s}}$ is the tangential stress due to a spatial variation of the interfacial tension.
 When this variation is due to a radially inhomogeneous heating, it is usually called thermocapillary stress.\\
 The variation of interfacial tension can be induced by the local heating and by the migration of surface active surfactants molecules at the interface.
 Generally, when there is no surfactants, $\partial \sigma /\partial T<0$ and interfacial tension is smaller at the hot spot
 but when the interface is charged with surfactants, the effective change of interfacial tension, involving coupling between temperature and concentration, can make interfacial tension larger at the hot spot.
In the present investigation, we consider the effective change of interfacial tension $\partial \sigma /\partial T$ due to both the heating and the solutal effect.
  More details on the change of behavior of interfacial tension with temperature, due to the presence of surfactants, can be found in the following references: \cite{rsv12}, \cite{khattari02}.\\
 The thermocapillary stress induces local flows on both sides of the interface which lead to its deformation.\\
Considering on the one hand small deformation amplitude of height $h(r)$ (i.e. $\partial h/\partial r<<1$), we can write the tangential stress condition as\\
\begin{equation}
 \eta_1\frac{\partial u_{1,r}}{\partial z}|_{z=h}-\eta_2\frac{\partial u_{2,r}}{\partial z}|_{z=h}
=\frac{\partial \sigma}{\partial s}\backsimeq \frac{\partial \sigma}{\partial r}=
  \frac{\partial \sigma}{\partial T} \frac{\partial T}{\partial r},\label{tang}
\end{equation}
where $T(r)$ is the temperature distribution at the interface due to laser heating.
On the other hand, assuming a Gaussian axisymmetric temperature distribution such as
\begin{equation}
T(r)=\Delta T_0e^{-r^ 2/w^2},
\end{equation}
This expression of $T(r)$ allows us to solve the hydrodynamics problem using the Fourier-Bessel transform defined such as
\begin{equation}
h(r)=\int_0^{\infty} h^k(k)J_0(kr)kdk\label{fourier-h}
\end{equation}
where $h^k$ is the Fourier-Bessel transform of $h(r)$, $k$ the reciprocal variable, and $J_0(x)$ the 0-order Bessel $J$ function.
We deduce
\begin{equation}
T(r)=\int_0^{\infty}T^k(k)J_0(kr)kdk,
\end{equation}
with 
\begin{equation}
T^k(k)=\Delta T_0\frac{w^2}{2}e^{-\frac{w^ 2 k^ 2}{4}} \label{temp}
\end{equation}
As we have $\frac{dJ_0(kr)}{dr}=-kJ_1(kr)$ where $J_1(x)$ is the 1-order Bessel $J$ function, we finally find
\begin{equation}
\frac{\partial T(r)}{\partial r}=-\int_0^{\infty}T^ k(k)J_1(kr)k^ 2dk
\end{equation}
Therefore, we find from equation (\ref{tang}) that $u_{i,r}(r,z)$ necessarily takes the form\\
\begin{equation}
u_{i,r}(r,z)=\int_0^{\infty}u^k_{i,r}(z)J_1(kr)kdk\label{urk}
\end{equation}
Moreover as the mass conservation condition (equation (\ref{mass})) can be written as\\
\begin{equation}
 \frac{1}{r}\frac{\partial (r u_{i,r})}{\partial r}+\frac{\partial u_{i,z}}{\partial z}=0\label{cont}
\end{equation}
it yields
\begin{equation}
 u_{i,z}(r,z)=\int_0^{\infty}u^k_{i,z}(z)J_0(kr)kdk\label{uzk}
\end{equation}
and the following relation between axial and radial velocities\\
\begin{equation}
 \frac{du^k_{i,z}(z) }{d z}=-ku^k_{i,r}(z) \label{rel}
\end{equation}
From Equations (\ref{urk}), (\ref{uzk}) and (\ref{rel}) we can now calculate the velocity field due to localized laser heating.
\subsection{Velocity field}
As far as one considers small interface deformation amplitude, the velocity field can be evaluated for 
the initially flat interface $h(r)=0$.\\
The stokes equations (equation (\ref{stokes})) can be written as follows in each liquid layer $i=1,2$\\
\begin{equation}
 -\frac{\partial p_i}{\partial r}+\eta_i\left(\frac{\partial^ 2}{\partial r^ 2}
+\frac{1}{r}\frac{\partial}{\partial r}-\frac{1}{r^ 2}+\frac{\partial^ 2}{\partial z^ 2}\right)u_{i,r}(r,z)=0\label{pr}
\end{equation}
\begin{equation}
 -\frac{\partial p_i}{\partial z}+\eta_i\left(\frac{\partial^ 2}{\partial r^ 2}
+\frac{1}{r}\frac{\partial}{\partial r}+\frac{\partial^ 2}{\partial z^ 2}\right)u_{i,z}(r,z)=0 \label{pz}
\end{equation}
Remembering the following properties of the Bessel functions\\
\begin{equation}
\frac{d^2J_1(kr)}{dr^ 2}+\frac{1}{r}\frac{dJ_1(kr)}{dr}-\frac{1}{r^ 2}J_1(kr)=-k^ 2J_1(kr)
\end{equation}
\begin{equation}
\frac{d^2J_0(kr)}{dr^ 2}+\frac{1}{r}\frac{dJ_0(kr)}{dr}=-k^ 2J_0(kr)
\end{equation}
we rewrite Equations (\ref{pr}) and (\ref{pz}) as\\
\begin{equation}
 -\frac{\partial p_i}{\partial r}+\int_0^{\infty}\eta_i\left(\frac{\partial^ 2}{\partial z^ 2}-k^ 2\right)u^ k_{i,r}(r)J_1(kr)kdk=0
\end{equation}
\begin{equation}
 -\frac{\partial p_i}{\partial z}+\int_0^{\infty}\eta_i\left(\frac{\partial^ 2}{\partial z^ 2}-k^ 2\right)u^ k_{i,z}(r)J_0(kr)kdk=0
\end{equation}
Eliminating the pressure and using equation (\ref{rel}), we finally get\\
\begin{equation}
 \left(\frac{\partial^ 2}{\partial z^ 2}-k^ 2\right)^ 2u^ k_{i,z}(z)=0
\end{equation}
This $4^{th}$ order linear equation of order $4$ defines the eigen modes of the axial velocity.\\
Its general solution is of the form\\
\begin{equation}
 u^ k_{i,z}(z)=A_ie^{kz}+B_ie^{-kz}+C_i\frac{z}{H_i}e^{kz}+D_i\frac{z}{H_i}e^{-kz}
\end{equation}
Using equation (\ref{rel}), the solution for the radial velocity is written\\
\begin{equation}
 u^ k_{i,r}(z)=-A_ie^{kz}+B_ie^{-kz}-\frac{C_i}{kH_i}(kz+1)e^{kz}-\frac{D_i}{kH_i}(-kz+1)ke^{-kz}
\end{equation}
where $A_i$, $B_i$, $C_i$ and $D_i$ are constants to be determined from boundary conditions.
We thus have eight unknowns to integrally solve the velocity field. The eight boundary conditions are:\\
Vanishing axial velocities at the interface ($z=0$)
\begin{equation}
u^k_{1,z}(z=0)=u^k_{2,z}(z=0)=0~~~~~B.C.1,2 \label{bc12}
\end{equation}
Continuity of the radial velocity at the interface ($z=0$)
\begin{equation}
 u^k_{1,r}(z=0)= u^k_{2,r}(z=0)~~~~~~B.C.3 \label{bc3}
\end{equation}
Vanishing velocities at the top $z=H_2$ and bottom $z=-H_1$ of the container (except for the symmetric configuration (b), where it is $\frac{d u^k_{1,r}}{d z}$ which is null at $z=-H_1$) 
\begin{equation}
 u^k_{1,z}(z=-H_1)=0~~~~~B.C.4 \label{bc4}
\end{equation}
\begin{equation}
 u^k_{1,r}(z=-H_1)=0~~~~~B.C.5a~~~or~~ \frac{d u^k_{1,r}}{d z}(z=-H_1)=0~~~~~B.C.5b\label{bc5}
\end{equation}
\begin{equation}
 u^k_{2,z}(z=H_2)=0~~~~~~B.C.6\label{bc6}
\end{equation}
\begin{equation}
 u^k_{2,r}(z=H_2)=0~~~~~~B.C.7\label{bc7}
\end{equation}
and finally the tangential stress jump at the interface (equation (\ref{tang})) for $h\simeq 0$
\begin{equation}
 \eta_1\frac{\partial u_{1,r}}{\partial z}|_{z=0}-\eta_2
\frac{\partial u_{2,r}}{\partial z}|_{z=0}= \frac{\partial \sigma}{\partial T} \frac{\partial T}{\partial r}~~~~~~B.C.8
\end{equation}
B.C.1 and 2 yield:\\
\begin{equation}
A_1=-B_1~~,~~A_2=-B_2~~~~~B.C.1,2 
\end{equation}
We can thus simplify notations and keep only 6 unknowns ($A_1,A_2,C_1,C_2,D_1,D_2$).\\
B.C.3 yield:\\
\begin{equation}
-2A_1-\frac{1}{k}\frac{C_1}{H_1}-\frac{1}{k}\frac{D_1}{H_1}=-2A_2-\frac{1}{k}\frac{C_2}{H_2}-\frac{1}{k}\frac{D_2}{H_2}~~B.C.3
\end{equation}
B.C.4 and 6 yield\\
\begin{equation}
A_1(e^{-kH_1}-e^{kH_1})-C_1e^{-kH_1}-D_1e^{kH_1}=0~~B.C.4
\end{equation}
\begin{equation}
A_2(e^{kH_2}-e^{-kH_2})+C_2e^{kH_2}+D_2e^{-kH_2}=0~~B.C.6 
\end{equation}
B.C.5 yield\\
\begin{equation}
-A_1(e^{-kH_1}+e^{kH_1})-\frac{1}{k}\frac{C_1}{H_1}(-kH_1+1)e^{-kH_1}-\frac{1}{k}\frac{D_1}{H_1}(kH_1+1)e^{kH_1}=0~~B.C.5a
\end{equation}
or 
\begin{equation}
A_1k(e^{-kH_1}-e^{kH_1})+\frac{C_1}{H_1}(-kH_1+2)e^{-kH_1}-\frac{D_1}{H_1}(kH_1+2)e^{kH_1}=0~~B.C.5b
\end{equation}
and B.C.7 leads to
\begin{equation}
-A_2(e^{kH_2}+e^{-kH_2})-\frac{1}{k}\frac{C_2}{H_2}(kH_2+1)e^{kH_2}-\frac{1}{k}\frac{D_2}{H_2}(-kH_2+1)e^{-kH_2}=0~~B.C.7
\end{equation}
Finally B.C.8 leads to
\begin{equation}
2\frac{\eta_2}{H_2}(C_2-D_2)-2\frac{\eta_1}{H_1}(C_1-D_1)=-\frac{\partial \sigma}{\partial T}kT^ k(k)~~B.C.8
\end{equation}
Boundary conditions $1-8$ entirely determine the velocity field produced by a localized laser heating and allow to 
calculate the resulting pressure field.
Note that when $H_1=H_2$, we have $u_r(-z)=u_r(z)$ and $u_z(-z)=-u_z(z)$ whatever $\eta_1/\eta_2$.
\subsection{Pressure field}
The pressure gradient along the $z$-axis (equation (\ref{pz})) can be written as\\
\begin{equation}
 \frac{\partial p_i}{\partial z}=\int_0^{\infty}\eta_i\left(\frac{\partial^ 2}{\partial z^ 2}-k^ 2\right)u^ k_{i,z}(z)J_0(kr)kdk
\end{equation}
Integration allows to determine the pressure in fluid $1$ and $2$ by specifying a reference pressure 
$p_{H_2}$ at $z=H_2$ 
\begin{equation}
 p_2(r,z)=\int_0^{\infty}\left(p_{H_2}+2\frac{\eta_2}{H_2}(C_2e^{kz}+D_2e^{-kz})\right)J_0(kr)kdk
\end{equation}
and
\begin{equation}
 p_1(r,z)=\int_0^{\infty}\left(p_{H_2}+2\frac{\eta_1}{H_1}(C_1e^{kz}+D_1e^{-kz})\right)J_0(kr)kdk
\end{equation}
which finally leads to
\begin{equation}
p_2(r,0)-p_1(r,0)=\int_0^{\infty}\left(2\frac{\eta_2}{H_2}(C_2+D_2)-2\frac{\eta_1}{H_1}(C_1+D_1)\right)J_0(kr)kdk
\end{equation}

\subsection{Interface deflection}
Knowing velocities and pressure fields, the shape of the interface can be deduced from the normal stress equation
 (equation (\ref{stressjump})) which can be re-written as
\begin{equation}
 \left(-p_1(r,z)+2\eta_1\frac{\partial u_{1,z}  }{\partial z }\right)_{z=h}-\left(-p_2(r,z)
+2\eta_2\frac{\partial u_{2,z}}{\partial z }\right)_{z=h}=\sigma \Delta_r h- (\rho_1-\rho_2)g h \label{def}
\end{equation}
Considering as previously the small deformation amplitude case, the left hand side of
 equation (\ref{def}) can be approximated by its value at $z=0$. It yields \\
\begin{equation}
\int_0^{\infty}4(\eta_1A_1-\eta_2A_2)kJ_0(kr)kdk=\sigma \Delta_r h- (\rho_1-\rho_2)g h
\end{equation}
Using the definition of the Fourier-Bessel transform (Equation (\ref{fourier-h}))and the relation $\Delta_r h=-\int_0^{\infty}k^ 2h^ k(k)J_0(kr)kdk$, where $\Delta_r$ denotes the radial part 
of the Laplacian operator, Equation ($\ref{def}$) becomes\\
\begin{equation}
4(\eta_2A_2-\eta_1A_1)k=(\sigma k^ 2 + (\rho_1-\rho_2)g) h^k(k)
\end{equation}
which leads to the expression of the induced deformation of the interface
\begin{equation}
h(r)= 4\int_0^\infty \frac{\eta_2A_2(k)-\eta_1 A_1(k)}{\sigma  k^ 2 + (\rho_1-\rho_2)g} J_0(kr)k^2dk 
\end{equation}
Defining the gravitational Bond number as $Bo=\frac{(\rho_1-\rho_2)g w^2}{\sigma}$, we finally find \\
\begin{equation}
h(r)= 4w\frac{\eta_2}{\sigma}\int_0^\infty \frac{A_2(k)-\frac{\eta_1}{\eta_2} A_1(k)}{w^2k^ 2 + Bo} J_0(kr)wk^ 2dk 
\end{equation}
In the following, we use the heating length $w\sim10-100\mu m$ as a reference length, $u_0=\frac{1}{\eta_1+\eta_2}|\frac{\partial\sigma}{\partial T}|\Delta T_0$ as a reference velocity
 and we define the thermal parameter $\alpha$ such as\\
\begin{equation}
\alpha=\frac{\partial \sigma}{\partial T}\frac{\Delta T_0}{\sigma}
\end{equation}
Assuming $\eta_i\sim  1- 100mPa.s$, $\sigma \sim 10^ {-3} N m^ {-1}$, $\Delta T_0\sim 10K$ and $\frac{\partial\sigma}{\partial T} \sim 10^ {-4}Nm^{-1}K^{-1}$,
 we find $\alpha \sim 1$ and $u_0\sim 5-500mm/s$. For all the calculations, we set $Bo=0.05$.\\
In the next section we present the velocity field and the interface deformation predicted analytically for two liquid layers separated by a flat interface (configuration (a)).
We investigate the influence of the layer thickness to the heating spot ratio $H_1/w$, the layer ratio $H_1/H_2$ and the viscosity ratio $\eta_2/\eta_1$ on both thermocapillary flows and interface deformation.
\section{Results and discussion}

\subsection{Double layer system}
\begin{figure}[h!!!]
\begin{center}
 \includegraphics[scale=0.5]{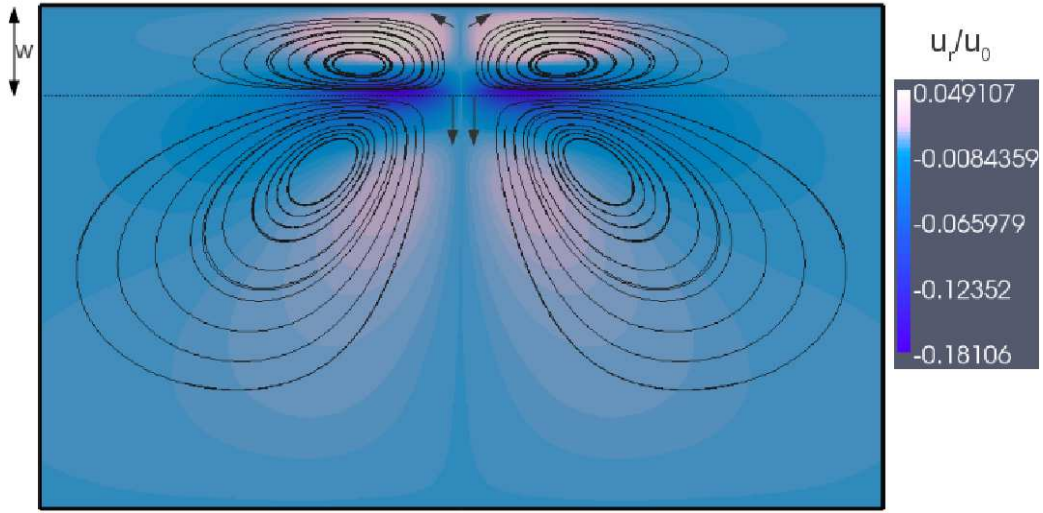}
\caption{Steady flow pattern in a double layer configuration for $H_1=5H_2=5w$, $\alpha=1$ and $\eta_2=\eta_1$.}\label{chp_vit_mur}
\end{center}
\end{figure}
\begin{figure}[h!!!]
\begin{center}
 \includegraphics[scale=0.6]{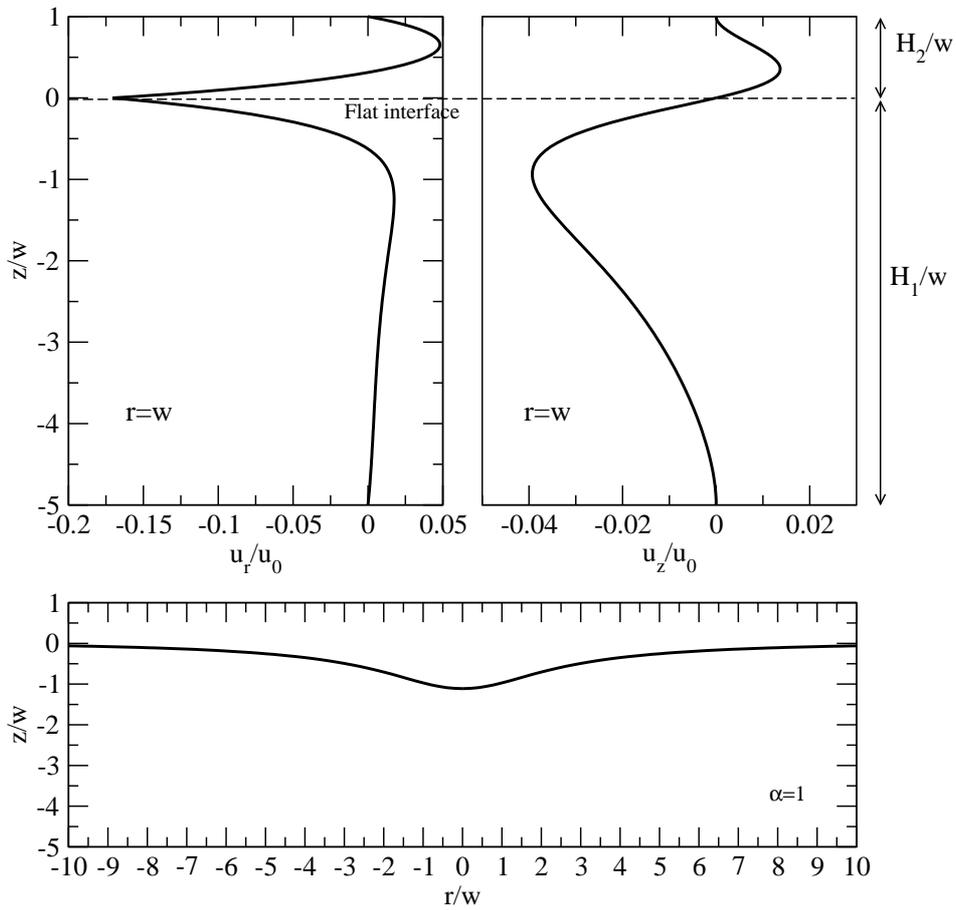}
\caption{Reduced radial velocity $u_r/u_0$ (Top left), reduced axial velocity $u_z/u_0$ (Top right) and interface deformation (Bottom) for
$H_1=5H_2=5w$, $\alpha=1$ and $\eta_2=\eta_1$.}\label{vit_h_mur}
\end{center}
\end{figure}
\subsubsection{Flow pattern}
An example of velocity field produced by the laser heating at the interface of a double layer configuration is reported in figure \ref{chp_vit_mur} for the case $\alpha>1$.
We observe the development of two contrarotative toroidal eddies induced by the tangential stress at the interface and bounded vertically by the interface and the walls.

The reduced radial and axial velocity, respectively $u_r/u_0$ and $u_z/u_0$, are reported in figure \ref{vit_h_mur} as a function of $z$.
We can notice that the maximum radial velocity is located at the interface ($z=0$) and that the axial velocity is larger in the bottom layer which is thicker.
Figure \ref{vit_h_mur} also shows that the deformation of the interface is directed towards the thicker layer when fluid viscosities are the same. In the case $\alpha<0$, flows 
and interface deformation are reverted.\\

\subsubsection{Influence of the heating length}
\begin{figure}[h!!!]
\begin{center}
 \includegraphics[scale=0.6]{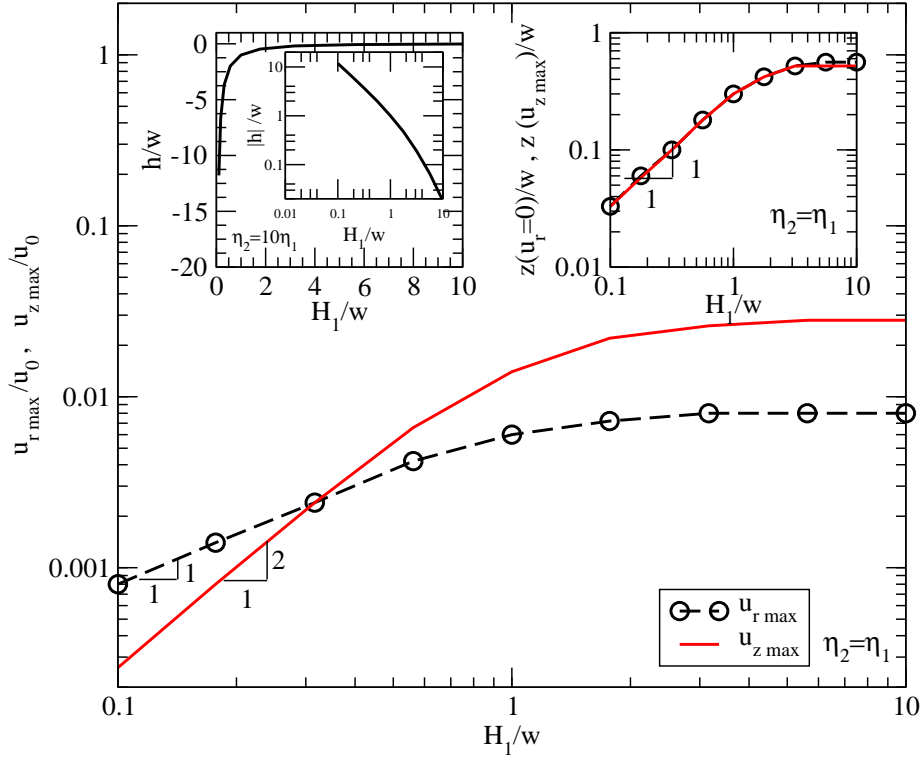}
\caption{Variations of the radial (continuous) and the axial (dashed with symbols) velocities versus the fluid layer reduced thicknesses $H_1/w$ ($H_1=H_2$ and $r=0.1w$).
 The right inset shows the reduced location $z_0/w$ of null radial velocity and maximum axial velocity ($z_0=z(u_r=0)=z(u_{z max})$). $\alpha=1$ and  $\eta_2=\eta_1$.
The left inset shows the reduced deformation height $h/w$ as a function of $H_1/w$ for $\eta_2=10\eta_1$. }\label{H-w}
\end{center}
\end{figure}
Figure \ref{H-w} shows that the maximum radial and axial velocity are increasing functions of $H_1/w$ for $\eta_2=\eta_1$. Note that when $H_1=H_2$ and $\eta_1=\eta_2$,
 there is no deformation of the interface because the resulting thermocapillary stress is null. We observe that when $H_1<<w$, both components depend on $H_1/w$, as the radial velocity $u_r$ scales like $u_0 H_1/w$ and the axial velocity $u_z$ scales like $u_0 (H_1/w)^2$.
However, when $H_1>>w$ both radial and axial components saturate and scale like $u_0$, the characteristic thermocapillary velocity.\\
The characteristic lengths of the flow pattern are reported in the right inset of figure \ref{H-w}. We notice that the largest amplitude of the axial velocity always coincides with a null radial velocity.
 The location of this characteristic point of the flow $z_0$ scales like $H_1$ when $H_1<<w$ as the liquid is bounded by a wall while it scales like $w$ when $w<<H_1$ as the flow magnitude significantly decreases beyond the heating spot size $w$.
The left inset of figure \ref{H-w} shows the variation of the interface deformation amplitude as a function of $H_1/w$ when the liquid viscosities are contrasted ($\eta_2=10 \eta_1$).
We can first notice that when $\alpha>0$ and $H_1=H_2$ the deformation is always directed towards the less viscous fluid. Moreover, the deformation amplitude decreases when increasing the layers thickness $H_1$ so that $H_1>>w$ leads to $h\rightarrow0$.
Even though not strictly quantitative we can approximate this decrease so that $|h|/w \sim w/H_1$ when $H_1<<w$.\\

\subsubsection{Influence of the liquid layer thickness ratio}
\begin{figure}[h!!!]
\begin{center}
 \includegraphics[scale=0.5]{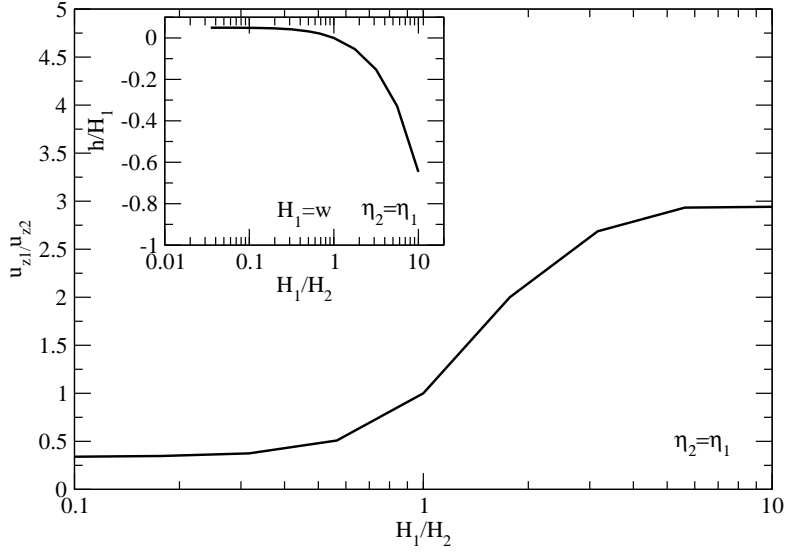}
\caption{Ratio of maximum axial velocities in each fluid $u_{z1}/u_{z2}$ as a function of $H_1/H_2$ in a double layer system.
 Inset shows the reduced deformation height $h/H_1$ as a function of $H_1/H_2$. $H_1=w$, $r=0.1w$ and $\eta_2=\eta_1$.}\label{H1-H2}
\end{center}
\end{figure}
Figure \ref{H1-H2} shows the variation of the maximum axial velocity ratio $u_{z1}/u_{z2}$ as a function of $H_1/H_2$. We chose $H_1=w$ to compare one of the heights to the length scale
of the thermal forcing.
 We first observe a saturation of $u_{z_1}/u_{z2}$ when $H_1>>H_2$ or $H_2>>H_1$ and that the axial velocity is always
 larger in the thicker layer. 
 The inset of figure \ref{H1-H2} shows that when $\eta_1=\eta_2$, the deformation is directed towards the thicker layer. 
When $H_1<<H_2$ (i.e. the heating characteristic length $w=H_1$ is much smaller than $H_2$), $h$ only depends on $w$, while in the opposite case, when $w>>H_2$, the reduced deformation amplitude $h/w$ strongly depends on $H_1/H_2=w/H_2$.
At the same time, the deformation amplitude $h$ increases as $H_2$ decreases as previously shown in figure \ref{H-w}.\\

\subsubsection{Influence of the viscosity ratio}
\begin{figure}[h!!!]
\begin{center}
 \includegraphics[scale=0.5]{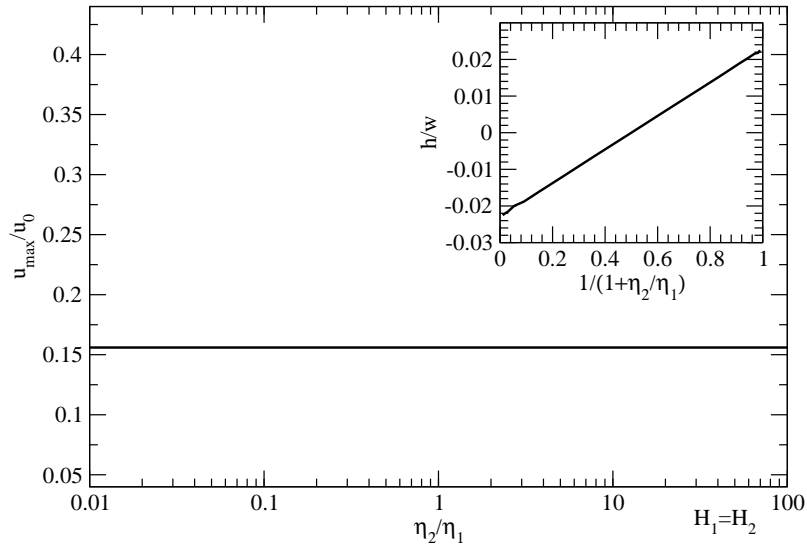}
\caption{Reduced maximum velocity $u_{max}/u_0$ as a function of viscosity ratio $\eta_2/\eta_1$ with $\eta_1=cste$.
 The inset shows the reduced deformation height $h/w$ as a function of $1/(1+\eta_2/\eta_1)$. $H_1=H_2$.}\label{eta2-eta1}
\end{center}
\end{figure}

An interesting feature of the thermocapillary flow in our configuration is the symmetrical properties of the velocities when $H_1=H_2$.
 In that case, we have $u_r(-z)=u_r(z)$ and $u_z(-z)=-u_z(z)$ for any value of $\eta_2/\eta_1$. Figure \ref{eta2-eta1} shows that $u_{max} \sim u_0$ 
and therefore $u_{max} \sim |\frac{\partial\sigma}{\partial T}|\Delta T_0/(2<\eta>)$.
This results can be retrieved using equation (\ref{tang}) and scaling arguments for $H_1>>w$. In this case $-\partial u_{r1} /\partial z \sim  \partial u_{r2} /\partial z \sim u_{max} / w$
 and $-\partial \sigma / \partial s \sim  \frac{\partial\sigma}{\partial T}\Delta T_0/w$.
Similarly, using scaling arguments and equation (\ref{def}), we find that $\Delta_r h \sim h/w^2 \sim 2(\eta_1-\eta_2) u_{max} /(w\sigma)$ and therefore 
$h/w \sim \alpha (\eta_1-\eta_2)/<\eta>$ as illustrated in the inset of figure \ref{eta2-eta1}.\\
Many features of the flows and interface deformation induced by the laser heating when $\partial \sigma / \partial T>0$ can be concluded from this section.
First, the flow magnitude increases with the layer thickness to the heating length ratio $H_1/w$ when $H_1<<w$ before reaching a saturation when $H_1>>w$.
When viscosities are equal in both layers, the velocity is larger in the thicker one and the interface deformation
 is always directed towards this same layer.\\
Moreover, when layer thicknesses are equal, $H_1=H_2$, the deformation is always directed towards
 the less viscous fluid. Its amplitude increases when decreasing the layers thicknesses as the vertical confinement increases the efficiency of the localized thermocapillary forcing.
Finally, keeping $H_1=H_2$, the dependence of the velocity magnitude and interface deformation can be predicted using simple scaling arguments that yield
 respectively to $u_{max} \sim |\frac{\partial\sigma}{\partial T}|\Delta T_0/(2<\eta>)$ and $h/w \sim \frac{\partial \sigma}{\partial T}\frac{\Delta T_0}{\sigma}(\eta_1-\eta_2)/<\eta>$.\\
In the next section, we investigate the laser heating of two fluid interface in a triple layer symmetric configuration where a thin film of liquid is bounded by two layers of a different liquid.
\subsection{Triple layer system}
\begin{figure}[h!!!]
\begin{center}
 \includegraphics[scale=0.37]{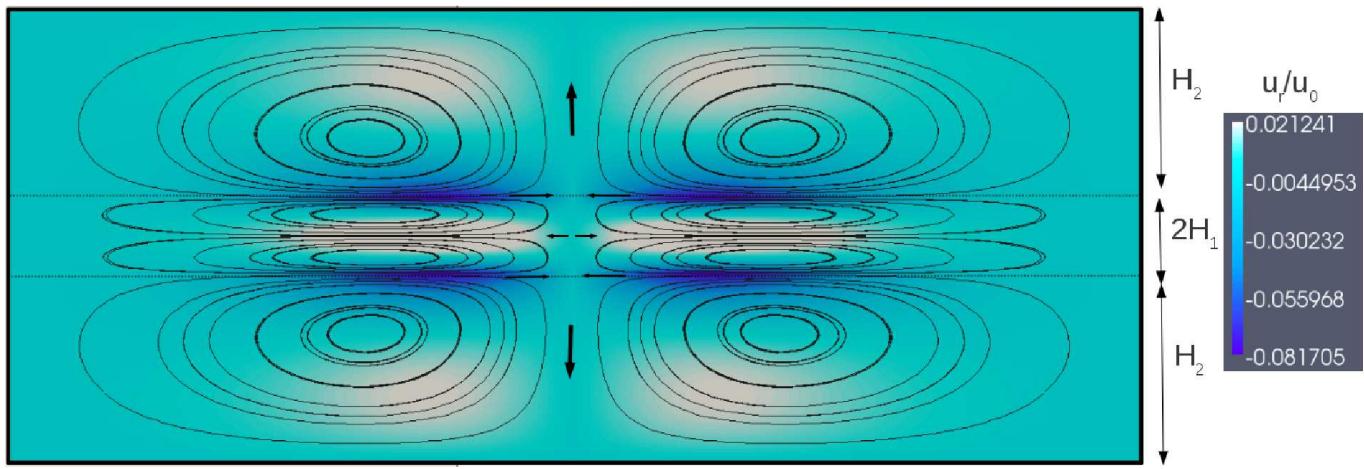}
\caption{Steady flow pattern in a triple layer configuration for $H_1=0.2w$, $H_2=w$, $\alpha=1$ and $\eta_2=\eta_1$.}\label{chp_vit_sym}
\end{center}
\end{figure}
The flow pattern showing the generation of eddies, in a system consisting of a thin film bounded by two external liquid layers of same thickness, is reported in figure \ref{chp_vit_sym}.
We can observe the production of two toroidal contrarotative eddies in the central film in addition to a toroidal eddy in each of the external liquid layers.
 The continuity of the radial velocity makes each pair of adjoining eddies contrarotative.\\
In this section, we investigate the influence of the film thickness to the external layer thickness ratio $H_1/H_2$ on the produced thermocapillary flows and interface deformations.\\
\subsubsection{Influence of the film thickness}
\begin{figure}[h!!!]
\begin{center}
 \includegraphics[scale=0.5]{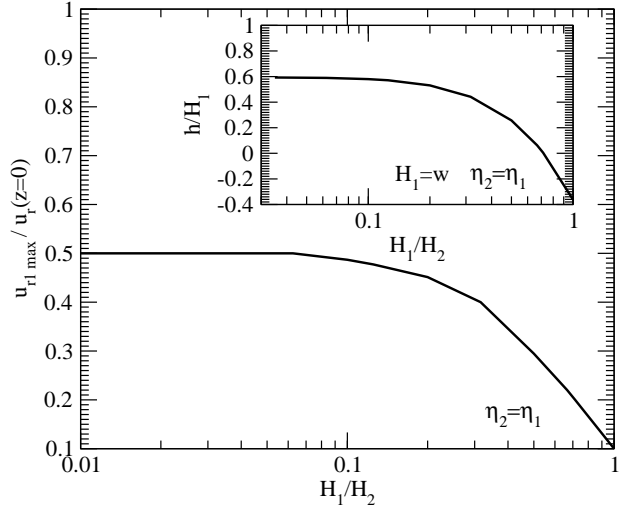}
\caption{Reduced maximum velocity $u_{r1 max}/u_r(z=0$) as a function of $H_1/H_2$ in a triple layer system. Inset shows the reduced deformation height $h/H_1$ as a function of $H_1/H_2$. $H_1=w$ and $\eta_2=\eta_1$.}\label{sym_H1-H2}
\end{center}
\end{figure}

\begin{figure}[h!!!]
\begin{center}
 \includegraphics[scale=0.5]{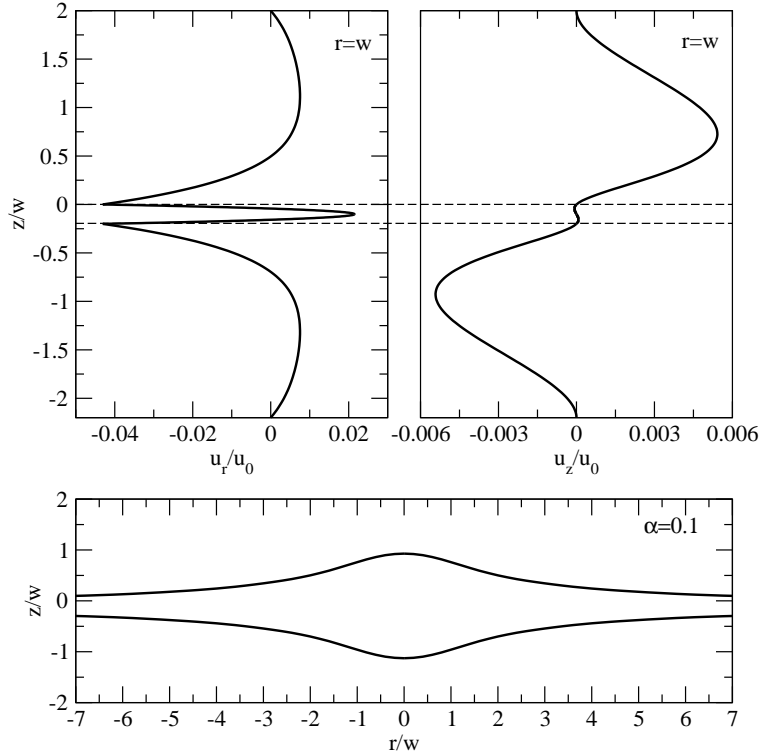}
\caption{Reduced radial velocity $u_r/u_0$ (Top left), reduced axial velocity $u_z/u_0$ (Top right) and interface deformation (Bottom) in a three-layer system for
$H_1=0.1w$, $H_2=2w$, $\alpha=0.1$ and $\eta_2=\eta_1$. }\label{dimple}
\end{center}
\end{figure}

\begin{figure}[h!!!]
\begin{center}
 \includegraphics[scale=0.5]{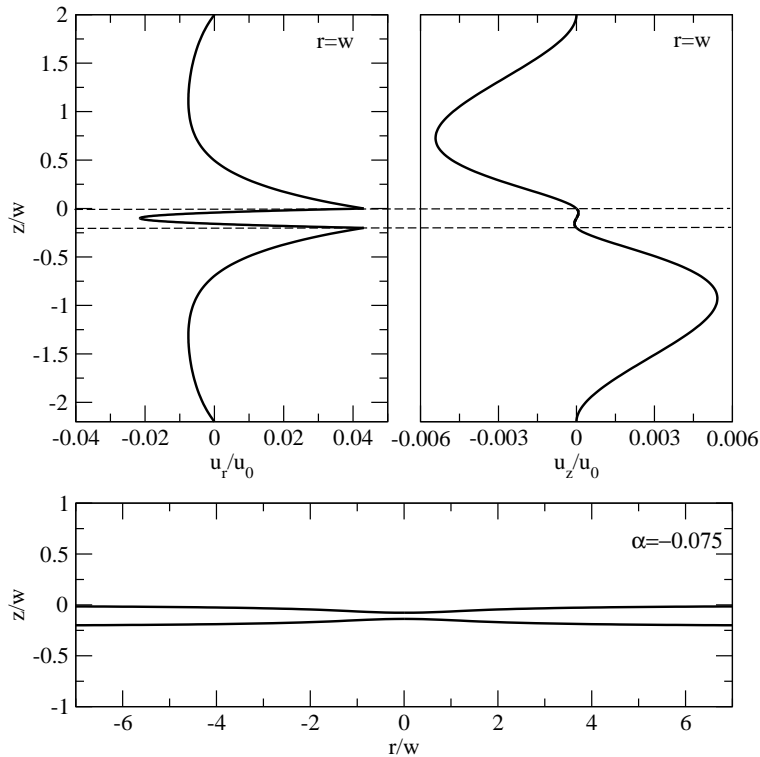}
\caption{Reduced radial velocity $u_r/u_0$ (Top left), reduced axial velocity $u_z/u_0$ (Top right) and interface deformation (Bottom) in a three-layer system for
$H_1=0.1w$, $H_2=2w$, $\alpha=-0.075$ and $\eta_2=\eta_1$.  }\label{nose}
\end{center}
\end{figure}

The maximum radial velocity reduced by its value at the interface as a function of $H_1/H_2$ is reported
 in figure \ref{sym_H1-H2} for equal viscosities ($\eta_1=\eta_2$).
We notice that the maximal radial velocity is constant for thin films ($H_1<<H_2$) as $u_{r max}=0.5 u_r(z=0)$
 while this velocity decreases when the film thickness becomes comparable to the external layer thickness.\\
The inset of figure \ref{sym_H1-H2} shows that the deformation is directed towards the external layers when $H_1<<H_2$ and its amplitude $h$ depends on $H_1$ while
its direction changes for $H_1/H_2 \simeq 0.7$.\\
It is interesting to study this system in both cases where the viscous
 tangential stress is directed towards the heating spot ($\partial \sigma / \partial r<0$) or in the opposite
 direction ($\partial \sigma / \partial r>0$).
In the first case, the deformation of the interface is directed towards the external layers inducing a
 dimple (figure \ref{dimple}) while in the second case, the deformation is directed towards the thin
 film inducing two noses (figure \ref{nose}).\\

\subsubsection{Inducing a dimple}
When the effective interfacial tension increases with temperature, the tangential stress is directed towards the heating spot. Assuming liquid layers
 of equal viscosities $\eta_1=\eta_2$, the deformation of the interface reported in figure \ref{dimple} is directed towards the external liquid layers of larger thicknesses.
This prediction has been observed experimentally in a system similar to our configuration. In a recent investigation, Dixit et al. \cite{Dixit09} used an infrared laser beam to 
induce the coalescence between two SDS-coated water drops in Decanol. The laser beam heats the rear of one of the drops and thermocapillary stresses force its migration towards the other drop.
The laser heating produces an accumulation of surfactant at the rear of the drop  \cite{baroud07} and therefore a deficit at the front which is in contact with the second non-heated drop.
This deficit increases interfacial tension and therefore creates tangential stresses directed towards the front of the drop near the second drop.
Coalescence would then occur at dimple edges location. Even though not strictly comparable, we believe that the dimple observed in this experiment
 is similar to that predicted by our analytical model and can be explained using the same arguments.
 The coalescence of liquid drops has also been observed experimentally in a different configuration \cite{baroud07b} where 
the heating laser intercepts the front and rear interfaces of water drops in contact, flowing in oil inside a microchannel, 
however the mechanism of the coalescence was not investigated in details. We believe that a dimple is formed in the thin film separating
 the drops before coalescence, similarly to the observation of Dixit et al. \cite{Dixit09}.\\

\subsubsection{Inducing noses}
In the case where the effective interfacial tension decreases with temperature,
 the tangential stress is directed from the hot region to the non-heated area of the interface.
 The direction of the flow is inverted and the deformations are directed towards the thin film inducing a nose (or hump) at each interface (see figure \ref{nose}).
 This could also lead to drop coalescence by forming a liquid bridge similar to the observation made in the experimental
 investigation led by Bremond et al. \cite{Bremond08} even though the mechanism at work here is quite different. 

\section{Conclusion}
We investigated the flows and the deformation of a liquid-liquid interface induced by the heating of a Gaussian laser beam in confined multi-layer systems.
A first part was dedicated to an interface separating two liquid layers in the case where the effective interfacial tension increases with temperature.
When the effective interfacial tension decreases with temperature, the direction of velocity and interface deformation is simply reverted.
The analytical resolution showed that the flow magnitude increases when increasing the thickness of the liquid layers
 before reaching a saturation when the thickness is much larger than the heating lengths scale. 
In the case of equal layer thicknesses, the deformation is always directed towards
 the less viscous fluid and its amplitude increases when decreasing the layers thicknesses. 
When the viscosities of the layers are equal, the velocity is always larger in the thicker layer and the deformation
 is always directed towards this same layer. In the special case where both viscosities and thicknesses are equal, there is no deformation
 of the interface.
A second part was dedicated to a thin film separating two layers of the same liquid.
We showed that decreasing the ratio of film thickness to the external layer thickness increases the deformation which is directed towards the external
 layer as far as the film is much thinner than the external layers.
Moreover, depending on the variation of interfacial tension with temperature, we showed that
 a dimple or a couple of opposite noses can be formed in the thin film,
 giving some insights for the explanation of drop coalescence observed in recent experiments.
 Although simple, this analytical resolution is predictive and therefore advances an interesting tool to explain and predict the features taking place in
 optofluidics experiments in which fluids are naturally confined.

\bibliographystyle{unsrt}

\end{document}